\documentclass[a4paper,12pt]{article} 
\usepackage{graphicx}
\usepackage{bm}         
\usepackage{amsmath}
\usepackage{amsfonts}
\usepackage{amssymb}
\newcommand{\vett}[1]{\mathbf{#1}}
\newcommand{\dert}{ \frac {\mathrm{d}~}{\mathrm{d}t} }

\newcommand{\dif}{\,\mathrm{d}}
\newcommand{\qj}{q^{(j)}}
\newcommand{\qk}{q^{(k)}}
\newcommand{\dqj}{ {\dot q}^{(j)} }
\newcommand{\ddqj}{ {\ddot q}^{(j)} }
\newcommand{\vqj}{ {\vett q}^{(j)} }
\newcommand{\vqk}{ {\vett q}^{(k)} }

\newcommand{\ret}{\mathrm{ret}}
\newcommand{\ug}{ \stackrel {\mathrm{def}} {=} }
\newcommand{\A}{\mathcal{A}}
\newcommand{\B}{\mathcal{B}}
\newcommand{\C}{\mathcal{C}}
\newcommand{\D}{\mathcal{D}}
\newcommand{\media}[2]{\frac {#1}{2\pi r} \oint #2  \dif l }	  
\newcommand{\lcor}{l}	  
\newcommand{\reali}{\mathbb{R}}
\newcommand{\sun}{\odot}
\newcommand{\pre}{Phys. Rev. E}
\newcommand{\apj}{ApJ}
\newcommand{\aap}{A\&A}
\newcommand{\pasj}{PASJ}
\newcommand{\physrep}{Phys. Rep.}
\newcommand{\aj}{AJ}
\begin{document}
   \title{Gravitational  effects of the  faraway matter on the rotation
  curves of spiral galaxies} 

   \author{A. Carati \thanks{  Dipartimento di Matematica, Universit\`a degli Studi Milano
     Via Saldini 50,   I-20133 Milano, Italy. Email:
     \texttt{andrea.carati@unimi.it} } }

   \date{\today}

\maketitle

 
  \begin{abstract}
    It was recently shown  
    that in cosmology the gravitational
    action of faraway matter has quite relevant effects, if retardation
    of the forces and discreteness of matter (with its spatial
    correlation) are taken into account.
    Indeed,  far matter was found
    to exert, on a test particle, a force per unit mass of the order of
    $0.2 \, cH_0$.
    It is shown here that such  a force 
    can account for the observed rotational velocity curves
    in spiral galaxies, if the force is assumed to be decorrelated beyond
    a sufficiently large distance, of the order of 1 kpc. In particular
    we fit the rotation curves of the galaxies NGC~3198, 
    NGC~2403, UGC~2885 and NGC~4725 without  any need of introducing dark
    matter at all. Two cases of galaxies presenting faster than keplerian 
    decay are also considered.
  \end{abstract}


%

\section{Introduction}

Perhaps, the actual problem of gravitation theory is to justify the
large speeds observed in objects of galactic or larger sizes.  This
problem became increasingly puzzling, starting from the end of the
years seventies. Indeed precise measurements (with a variety of
techniques) of the speed of gases in spiral galaxies made apparent
that the action of the galaxy on the gas, computed according to the
standard theory of gravitation, cannot account for the measured
velocities, if the gas is assumed to be a stable component of the
galaxy.  In the literature, while a lot of efforts are devoted to
computing the gravitational action of a galaxy on the nearby gas, no
attention is paid to the gravitational effects exerted by the other
galaxies, particularly the far away ones.  This attitude, of
neglecting the faraway matter, would indeed be correct if the other
galaxies were uniformly distributed. However, up to the present limit of
observations (at least 300 Mpc, according to Gabrielli et
al. \cite{gabrielli99, gabrielli}; Sylos Labini et al. \cite{sylos}),
it was suggested that the
galaxies may instead be 
distributed according to a very complex fractal structure.  It was
shown by Carati et al. (\cite{carati}) that the gravitational action of 
faraway matter computed according to the General Relativity
has quite relevant effects, if both the retardation of the
forces and the discreteness of matter (with its spatial correlation)
are taken into account. Indeed, the far matter was found to exert, on
a test particle, a force per unit mass of the order of $0.2\, cH_0$,
where $H_0$ is the Hubble constant, and $c$ the speed of light.  It
was also shown how the gravitational force due to the faraway matter
may account for the speed of the galaxies in clusters.
 
The aim of the present paper is to estimate the  effect 
that the faraway matter  produces on the rotation curves in spiral galaxies. 
We show that such an effect becomes very relevant if 
one assumes that  
the gravitational field due to the faraway matter is decorrelated beyond 
a sufficiently large distance (of the order of 1 kpc). 
So such a force goes unnoticed for objects of the solar system's size
(or smaller), 
while starts becoming important from objects of galactic 
size on, being the predominant one for  clusters of  galaxies and larger 
structures. In particular, we show that the rotation curves of four 
galaxies (with very different masses and sizes) can be fitted taking into 
account such a force, leading to reasonable values of the luminosity--mass 
ratio (see Table I of Section~4). Two galaxies presenting a faster than 
keplerian decay are also considered.

The paper is organized as follows. In Section~2, we recall the model that was 
introduced by Carati et al. (\cite{carati}) in order to deal with the 
gravitational effect of faraway matter. In Section~3 the 
effect on the rotation curves is evaluated, and in  Section~4  the fit 
to rotational curves for the galaxies  
NGC 3198, NGC 2403, NGC 2885, NGC 4725 is reported. Two cases of rotation 
curves presenting a faster than keplerian dacay are also briefly discussed.
Two Appendices, in which  some technical computations are reported,
complete the paper.  

\section{Definition of the model}
We briefly recall in this section the model that was  introduced by
Carati et al. (\cite{carati}) in order to estimate the gravitational 
action of far
away matter. To this end, 
first of all one has to estimate the gravitational field generated
by the distant galaxies (thought of as point masses) considered as sources
in  Einstein's equation, in the approximation in
which  their positions and velocities are assigned, in the
simplest way compatible with observational cosmology.  

To this end, the positions $\vett q_j$ of the galaxies
are considered as random variables, whose
statistical properties will be discussed later.
The velocity field of the sources is instead taken according 
to  Hubble's law, the peculiar velocities being  altogether neglected.
Namely, 
taking a locally Minkowskian coordinate system
centered about an arbitrary point, a particle with position vector
$\vett q$ is assumed to have a velocity
\begin{equation}\label{eq:hubble}
\dot{\vett q}=H_0 \, \vett q\ ,
\end{equation}
where, for the sake of simplicity of the model, the Hubble constant $H_0$
is assumed to be time independent.  It is easily established
that the considered chart has a local 
Hubble horizon $R_0=c/H_0$, where the galaxies' speed equals that of
light.  

The energy--momentum tensor $T^{\mu\nu}$ is then given by
\begin{equation}\label{eq:TEM}
T^{\mu\nu}=\sum_{j=1}^N \frac 1 {\sqrt g}\ \frac {M_j}{\gamma_j}\
\delta (\vett x-\vett q_j) \dot{\vett q}^\mu_j \dot{\vett q}^\nu_j\ ,
\end{equation}
where $N$ is the number of galaxies,
 $M_j$ and $\gamma_j$ are the mass and the Lorentz factor of the
$j$--th one, $g$ is the determinant of the metric tensor (which
is considered as an unknown of the problem), $\delta$ the Dirac delta
function, and the dot denotes derivative with respect to proper time
along the worldline of the source.  

The study of the solutions of Einstein's equations with the
energy--momentum tensor (\ref{eq:TEM}) as a source  is a
formidable task, and so the study of Carati et al. (\cite{carati})  was 
performed
within the limits of a perturbation approach,
considering the energy--momentum tensor $T^{\mu\nu}$ (\ref{eq:TEM})
 as a perturbation of the vacuum. Following the standard
procedure (see Einstein \cite{einstein} or Weinberg \cite{weinberg}), 
one then has to determine a
zero--th order solution (the vacuum solution), and solve the Einstein
equations, linearized about it.  The simplest consistent zero--th
order solution is the flat metric because,
coherently, the perturbation is then shown  to be small (at least if the
parameters of the model, such as the density, 
are chosen in agreement with the observations). 

Thus the metric tensor $g_{\mu \nu}$ is written as a perturbation of
the Minkowskian background $\eta_{\mu \nu}$, namely, as $g_{\mu
\nu}=\eta_{\mu \nu} +h_{\mu \nu}$, and it is well known that in the
linear approximation the perturbation $h_{\mu\nu}$  then has to satisfy
essentially the wave equation with $T^{\mu\nu}$ as a source. More
precisely, one gets
\begin{equation}\label{eq:EqEin}
\square \big[ h_{\mu\nu}-\frac 12 \eta_{\mu\nu}h\big]= -\frac{16\pi
  G}{c^4} T_{\mu\nu}\ ,
\end{equation}
where $G$ is the gravitational constant, $h$ the trace of
$h_{\mu\nu}$, and $\square=(1/c^2)\partial^2_t-\Delta_2$.  The
solutions are the well known retarded potentials
\begin{equation}\label{eq:RetPot}
h_{\mu\nu}=\frac {-2 G}{c^4 }\, \, \sum_{j=1}^N \frac {
M_j}{\gamma_j}\, \left. \frac {2\dqj_\mu \dqj_\nu -c^2\eta_{\mu\nu}}
{|\vett x-{\vett q}_j|}\right|_{t=t_{\ret}} \
\end{equation}
(with ${\vett q}^{(j)}\equiv {\vett q}_j$). 

In the present case in which we are concerned with the rotation curves 
of spiral
galaxies, one has also to take into account the contribution of local
matter, so that,  
in the spirit of the considered
approximation, the total metric tensor is  given by the sum
of that due to the local matter, and of that due to the distant
one. For the local metric
$h_{\mu\nu}^{\mathrm{loc}}$, one furthermore makes the Newtonian
approximation, i.e., all components are assumed to vanish, apart from
$h_{00}^{\mathrm{loc}}$, which  is set equal to the Newtonian potential
$V^{loc}(\vett x)/2c^2$ due to local masses. 
By local matter we mean the one
contained in the galaxy, so that we forget, in the rest of the paper, 
the possible contribution of the dark matter halo. 
The total metric is thus written as
$$
 g_{\mu\nu}=\eta_{\mu\nu} +  \frac {2V^{loc}(\vett x)}{c^2} \delta_{0\nu}
 \delta_{0\mu} + h_{\mu\nu} \ ,
$$
with $h_{\mu\nu}$ given by (\ref{eq:RetPot}).

So the equation of motion for a test particle  contains,
in addition to the  Newton force produced by the mass inside 
the galaxy, coming from the term $h_{\mu\nu}^{\mathrm{loc}}$,
also some terms coming from the field due to the distant matter. How large is
this additional term with respect to the local Newton force? The
answer depends,  
on the assumptions made on the positions $\vett q_j$ of the
constituents of the far away matter which appear in formula (\ref{eq:RetPot}).

Carati et al. (\cite{carati}) assumed the positions $\vett q_j$
 to be random variables, having a probability distribution
which is isotropic (rotationally invariant). 
A simple computation then shows that the mean force (per unit mass)
acting on a test particle vanishes. Thus the order of magnitude of the force
is given by its standard deviation. 

The estimate of the standard deviation requires an additional hypotesis
on the distribution of the quantities $\vett q_j$. If they are assumed
to be independent and  identically distributed, then the force turns out to
be negligible.  

On the other hand, it is known that the positions of
different galaxies are correlated, having a distribution
with a fractal character, and it has been suggested that this may
happen up a certain quite large distance (see  Sylos Labini et al. \cite{sylos};
Gabrielli et al. \cite{gabrielli,gabrielli99}). 
Carati et al. (\cite{carati}), assuming that
the distribution is fractal at all distance scales, 
show that the force per unit mass due to the distant galaxies is
of the order of $0.2\, cH_0$.

Now, as pointed out by Milgrom (\cite{milgrom,milgrom87}),
it is precisely when the acceleration of the stars is of the order of
$cH_0$  that the rotation
curves of the spiral galaxies depart from the behaviour expected from
the Newtonian force due to local visible matter. 
Our purpose is then to compute the effect of the far field, precisely
in such a region. It will be shown that such an the effect can be described as
corresponding to an
effective potential which acts on the stars by modifying the profile of
the rotation curves as to fit the actual observed one.

\section{The effect of far away matter on the rotation curves}

In this Section, we estimate the effect of the field produced
by the far away matter on the rotation curves. So, first of all, one
has to estimate the influence of the far field on the motion of
stars. 

In general, as there is no reason to expect that the far field
should be spherically symmetric, its first effect amounts to destroy the
circular orbits, from which every discussion about the rotation curves
usually starts. To tackle this poblem, one can imagine that the effect
of the
far field is not too large, and that the orbits are still close to
circular ones. In this case one can average the radial component of the
equation of motion along a circle, thus obtaining an effective radial
equation from which the angular velocity of the stars
as a function of the radius may be estimated.

Using cylindrical coordinates $(r,\theta,z)$, and considering only
motions which lie 
in the galactic plane $z=0$ (so that $\dot z=$$\ddot z=0$), the radial 
equation of motion turns out to be 
\begin{eqnarray}\label{eq:EqGeod} 
    2 \dert \Big(g_{rr} \dot r &+& g_{r\theta}\dot\theta + g_{r0}c \Big) 
    =    c^2  \partial_r g_{00} +
    \dot {\theta}^2 \partial_r  g_{\theta\theta}  
    + 2 c \dot\theta \partial_r g_{0\theta} \nonumber \\
        &+& 2 \dot r\dot\theta \partial_r g_{r\theta} + 2 c\dot
    r \partial_r g_{r0} + \dot r^2 \partial_r g_{rr}
    \ .
\end{eqnarray}
In particular, in terms of the cartesian components $h_{\mu\nu}$ 
given by (\ref{eq:RetPot}) and of the local potential
$h_{\mu\nu}^{\mathrm{loc}}$, the relevant components are given by the formulas
\begin{eqnarray}\label{eq:gcordcil}
    g_{rr} &=& r^2\big( 1+ h_{11}\cos^2 \theta +h_{22}\sin^2 \theta 
               + h_{12} \sin \theta \cos \theta  \big) \nonumber \\
    g_{r\theta} &=& r\big( ( h_{11} - h_{22} )\sin\theta \cos\theta
               +h_{12}\cos 2\theta \big) \ug r\D \nonumber \\
    g_{0r} &=& h_{01}\cos \theta + h_{02}\sin \theta  \nonumber \\
    g_{\theta\theta} &=& r^2\big(1+ h_{11}\sin^2 \theta +h_{22}\cos^2 \theta 
               + h_{12} \sin \theta \cos \theta  \big) \ug r^2(1+\A) \nonumber \\
    g_{0\theta} &=& r\big( h_{02}\cos \theta - h_{01}\sin \theta  \big)
               \ug r\B   \nonumber \\ 
    g_{00}  &=& 1+ h_{00} + \frac {2}{c^2} V^{loc}(r) \ .
\end{eqnarray}
The quantities $\A$, $\B$ and $\D$, implicitly defined in such formulas,
will be used later. In our setting, equation
(\ref{eq:EqGeod}) is a stochastic one, because of the stochastic
character of the terms $h_{\mu\nu}$, and it is very hard to be dealt with.

To reduce our problem to a tractable one, we make essentially two
hypotheses. First of all we suppose that it is meaningful to average
the equation with respect to $\theta$, i.e., we suppose that $\theta$
is a fast variable with respect to $r$, and consequently the derivative
$\dot r$ must be thought of as ``small''. Thus, in the averaged equation,
we will forget all terms proportional to $\dot r$.

Now, as the galaxies do not change their dimensions on the time scale of the
revolution of a typical star, one can readily give an estimate of the
term $ \mathrm{d} (g_{rr} \dot r )/\mathrm{d}t$
of (\ref{eq:EqGeod}), which turns out to essentially
vanish. Obviously, also the terms which are the 
derivatives with respect to $\theta$ of components of the
metric tensor will have a vanishing averages. One is therefore left with the
following equation in the unknown $v\ug r\dot\theta$:
\begin{eqnarray}\label{eq:Fond1}
    0 = &\partial&_r V^{loc} + \Big[ 1+\media{1}{(\A+r\partial_r
	\A)}\Big] \frac {v^2}{r} \nonumber \\
        &+& \Big[ \media{c}{(\B+r\partial_r\B)}  
         - \media{1}{\partial_t \D} \Big]\frac {v}{r} \nonumber \\
       &-& \media{c}{\partial_t g_{0r}} + \media{c^2}{\partial_r h_{00}} \ ,  
\end{eqnarray}
where use was made of expressions
(\ref{eq:gcordcil}) for the 
components of the metric tensor.
This equation has to be compared (see for example Bertin \cite{bertin})
with the familiar one
$$
v^2/r=-\partial_rV^{loc} \ ,
$$ 
which is  obtained  when the only acting force is assumed to be 
the Newtonian one due to local matter. 
So, at first sight, the structure of the equation seems to have changed
just a little bit: there appears a term which is linear in $v$, but this
modification turns out to be negligible. However, a more relevant contribution 
(in some hypotheses to be discussed in a moment) is given by the term
\begin{equation}\label{eq:Peff}
\media{c}{\partial_t g_{0r}} \ug  - \partial_r V^{eff}\ .
\end{equation}
In this respect, the faraway field acts as to produce a modification of the
local potential, giving rise to a sort of effective potential; 
this effect might also be mimicked by a local distribution of matter (of an
unknown type). 

The relevant difference between our formula (\ref{eq:Fond1}) and
the standard one, consists in the fact that in  (\ref{eq:Fond1})
the terms depending on the far fields have a stochastic character, so
that, as already pointed out, we cannot predict their exact values, 
but can only estimate their order of magnitude by
looking at their average and variance.

Notice that, as we are integrating along a circular
path, the magnitude of the random terms will depend on the correlation of
the field at different points of the path. Being interested only in
the variance, it will suffice to just consider  the two--point
correlations.

In principle such correlations should be
computed from the expression (\ref{eq:RetPot}), but at the moment we are
unable to do this in a consistent way. Therefore we make here our
second assumption, i.e., it will be assumed that the
correlations decay in an exponential way on a certain lenght scale
$\lcor$, namely, we assume
\begin{equation}\label{eq:corr1}
  \big\langle \partial_\sigma g_{\mu\nu}(x)\, \partial_\sigma g_{\mu\nu}(y)
  \big\rangle =   \exp\Big( -\frac {|x-y|}{\lcor} \Big) 
  \big\langle \big( \partial_\sigma g_{\mu\nu}(x) \big)^2 \big\rangle \ .
\end{equation}
Here $\lcor$ is considered as a free parameter, to be fitted by
comparison with the actual observations of galaxy rotation curves. If
one obtains consistent values of $\lcor$ for different galaxies, the ansatz
(\ref{eq:corr1}) might be considered as sound. If $\lcor$ is large, the
field can be considered as smooth, becoming more stochastic as $\lcor$
decreases. We will find for $\lcor$ a value of $1$ kpc, so that the
far field can 
be considered as stochastic for what concerns the motion of stars
in a galaxy, and smooth instead, for example for what concerns the motion of
planets around a star.

We recall that, via the Wiener--Khinchin formula (see Appendix~B),
one can obtain the correlation of a function, from the correlations
of its derivatives. In our case, we obtain
\begin{equation}\label{eq:corr2}
  \big\langle g_{\mu\nu}(x)\, g_{\mu\nu}(y)
  \big\rangle =   \frac {\lcor^3}{|x-y|}\Big( 1-e^{-\frac{|x-y|}{\lcor}}
  \big(  1+\frac {|x-y|}{2\lcor} 
 \big) \Big) \big\langle \big( \partial_\sigma g_{\mu\nu}(x) \big)^2
  \big\rangle \ . 
\end{equation}
Using (\ref{eq:corr1}) and (\ref{eq:corr2}), the
averages and variances of
all terms involved in equation  (\ref{eq:Fond1}) can be computed as
shown in Appendix~A. The results of such computations are the
following. Denote
\begin{eqnarray*}
    \A_1 &\ug& \media{1}{\A} \ ,\quad \A_2 \ug \media{1}{\partial_r \A}\ , \\
    \B_1 &\ug& \media{1}{\B} \ ,\quad \B_2 \ug \media{1}{\partial_r \B}\ , \\
    \C   &\ug& \media{c}{\partial_t g_{0r}} \ ,\quad 
    \D_1 \ug \media{1}{\partial_t \D} \ . 
\end{eqnarray*}
Then, for the averages and the variances  one finds the following results:
\begin{eqnarray}\label{eq:scarti}
    \langle \A_1 \rangle &=& -\frac 12 
                                                             \nonumber \\  
    \langle \A_2 \rangle &=& \ \ 0 \ , \quad \sigma^2_{\A_2} = \frac
    {\lcor}{\pi r} \langle (\partial_r \A )^2\rangle = \frac
    {\gamma_A}{\pi} \Big(\frac {H_0}{c}\Big)^2  \frac {\lcor}{r}
                                                             \nonumber \\
    \langle \B_1 \rangle &=& \ \ 0 \ , \quad \sigma^2_{\B_1} =\frac {\lcor^3}{r}
    \langle (\nabla \B )^2\rangle = \frac {3\gamma_B}{4\pi} \Big(\frac
	    {H_0}{c}\Big)^2  \lcor^2\Big(\frac {\lcor}{r}\Big)^{0.8} 
                                                             \nonumber \\
    \langle \B_2 \rangle &=& \ \ 0 \ , \quad \sigma^2_{\B_2} = \frac
    {\lcor}{\pi r}
    \langle (\partial_r \B )^2\rangle = \frac {\gamma_B}{4\pi} \Big(\frac
    {H_0}{c}\Big)^2  \frac {\lcor}{r} 
                                                             \nonumber \\ 
    \langle \C   \rangle &=& \ \ 0 \ , \quad \sigma^2_{\C} = 
    \frac {c^2\lcor}{\pi r} \langle (\partial_t g_{r0})^2 
    \rangle = 0.05 H_0^2 c^2 \frac {\lcor}{\pi r} 
                                                             \nonumber \\
    \langle \D_1 \rangle &=& \ \ 0 \ , \quad \sigma^2_{\D_1} = \frac {\lcor}{r}
    \langle (\partial_t g_{r\theta})^2\rangle = \frac 1{4\pi} \gamma_D H_0^2
    \frac {\lcor}{r}    
    \ ,
\end{eqnarray}
where the constants  are given by $\gamma_A\simeq 0.05$,
$\gamma_B\simeq 0.026$ and $\gamma_D\simeq 0.08$. 

Notice that in the first line the variance is not computed because,
being the average not vanishing, it is just the average which gives
the order of magnitude of the terms in equation (\ref{eq:Fond1}).
Using relations
(\ref{eq:scarti}), for  $r$ in the range of 
galactic distances, i.e., $r \le 30$ kpc, one checks that
all terms in (\ref{eq:Fond1}) are negligible, except for three of them: 
the local
Newtonian potential, the quadratic term in the tangential velocity
$v$, and the term  containing $\oint \partial_t g_{0r}\dif l/2\pi
  r$. Thus (\ref{eq:Fond1}) takes the simpler form 
\begin{equation}\label{eq:Fond2}
\frac 32 \frac {v^2}{r}= -\partial_r V^{loc}(r) - \partial_r V^{eff}
 \ , \quad \mbox{with }\  
 \partial_r V^{eff} \simeq 0.2 H_0c\sqrt{\frac{\lcor}{r}} \ .
\end{equation}
Here, the value indicated after the sign $\simeq$ is just the standard
deviation. Indeed, as explained in the
Introduction, we are dealing with random variables having zero mean,
so that their order of magnitude is given by their standard deviation.
Remember that the sign can be
positive or negative: if the term is positive  
it acts as a pressure and so it helps keeping the galaxy
stable, while the opposite occurs if it is negative. So one can
conjecture that the positive case occurs more frequently in
observations, but it would be possible in principle to observe also
cases in which the term has a negative value.
\begin{figure}
  \centering
\resizebox{\hsize}{!}{\includegraphics{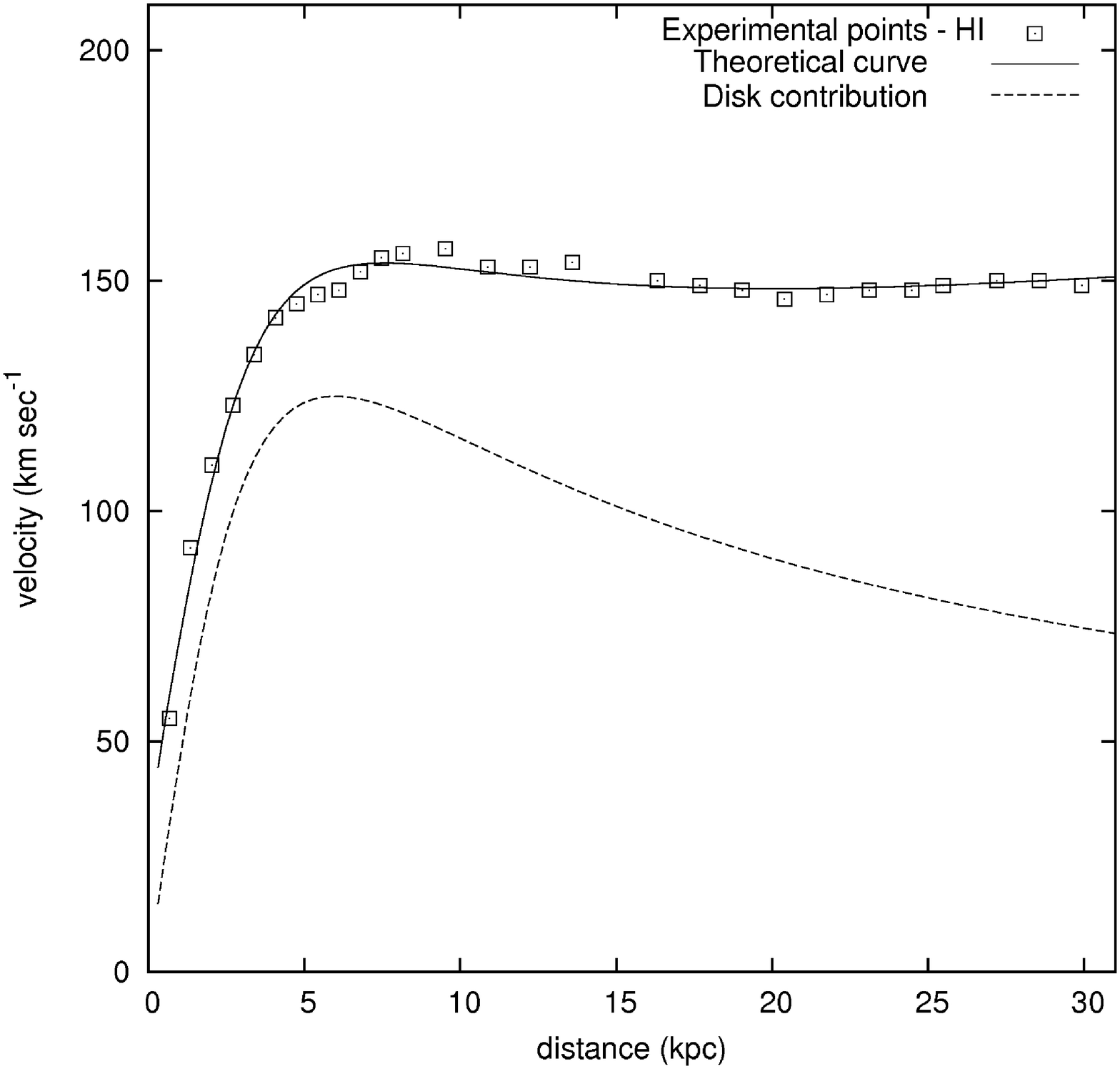}}
  \caption[Rotation curve for galaxy NGC 3198]{\label{fig1} Plot of
  the rotational velocity $v$ vrtsus 
   distance $r$ from the galactic center, for the
  galaxy NGC 3198. Squares ($\boxdot$) are the velocities determined from HI
  observations (see Bergman \cite{begeman89}), solid line is the
  theoretical curve (see 
  text). Dashed line is the contribution due to the local matter. 
  }
\end{figure}

We show in the next section that, if the term is positive, the
magnitude of the correction due to the far field
is able to flatten the rotation curve in the external region of
the galaxy.

\section{Fitting some observed rotation curves}
\begin{figure}
  \centering
\resizebox{\hsize}{!}{\includegraphics{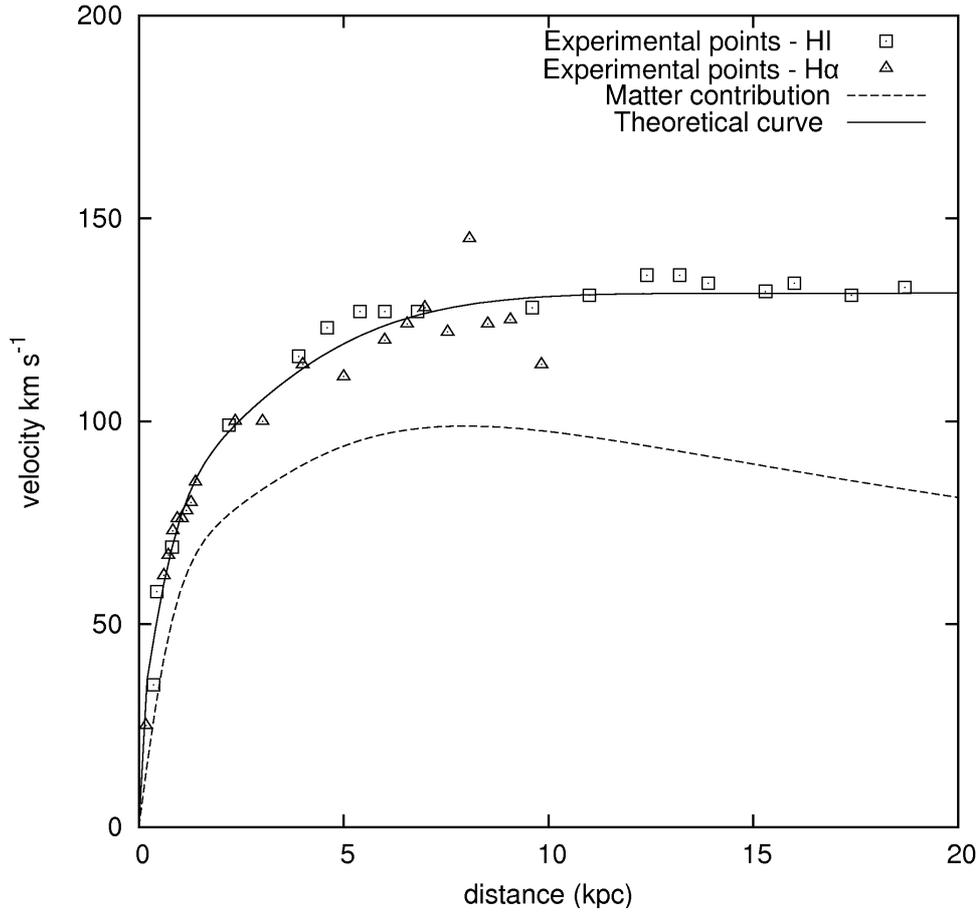}}
  \caption[Rotation curve for galaxy NGC 2403]{\label{fig2} Plot of
  the rotational velocity $v$ versus
   distance $r$ from the galactic center, for the
  galaxy NGC 2403. Squares ($\boxdot$) are the velocities determined from HI
  observations (see Bergman \cite{begeman87}), while triangles 
($\triangle \kern
  -1.4ex \cdot\; $)  
   are H$\alpha$ observations
  (see Blais-Ouellette et al. \cite{blais2004}).  
   The solid line is the theoretical curve (see
  text), while the dashed line is the contribution due to the local matter. 
  }
\end{figure}
\begin{figure}
  \centering
\resizebox{\hsize}{!}{\includegraphics{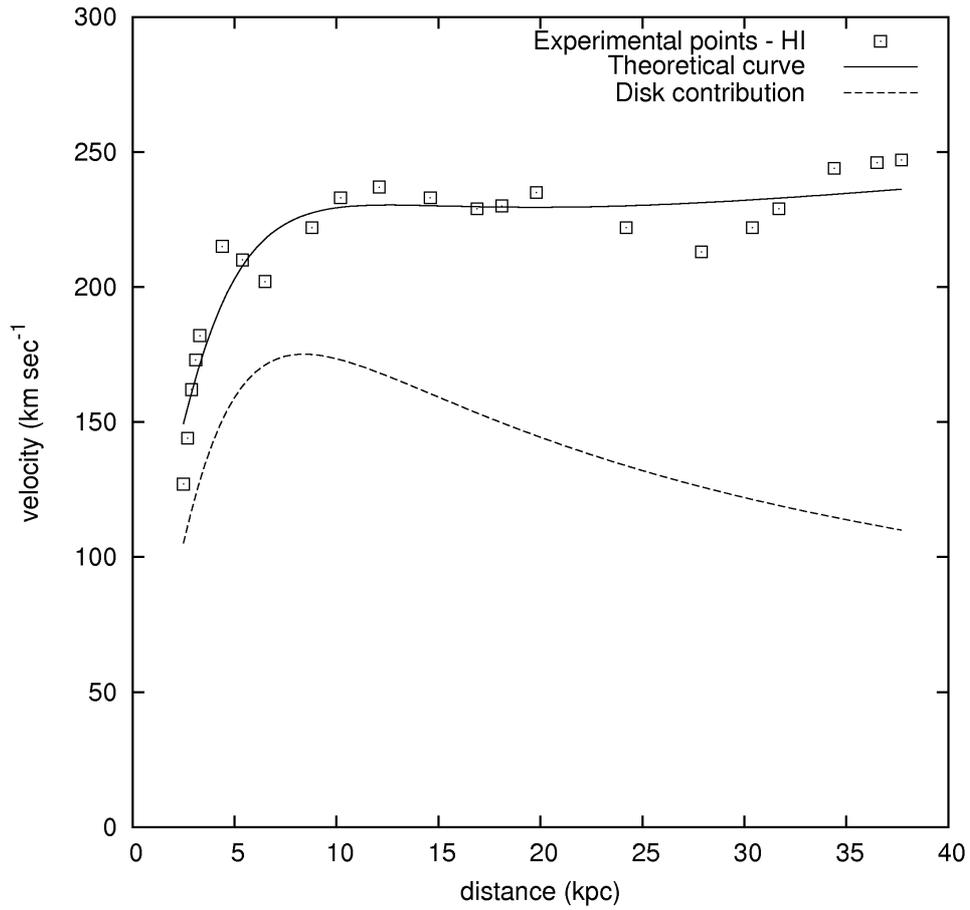}}
  \caption[Rotation curve for galaxy NGC 4725]{\label{fig4} Plot of
  the rotational velocity $v$ versus 
  distance $r$ from the galactic center, for the
  galaxy NGC 4725. Squares ($\boxdot$)  are the velocities determined from HI
  observations (see Wevers et al. \cite{wevers84}). The solid line 
  is the theoretical curve (see 
  text), while the dashed line is the contribution due to the local matter. 
  }
\end{figure}
\begin{figure}
  \centering
\resizebox{\hsize}{!}{\includegraphics{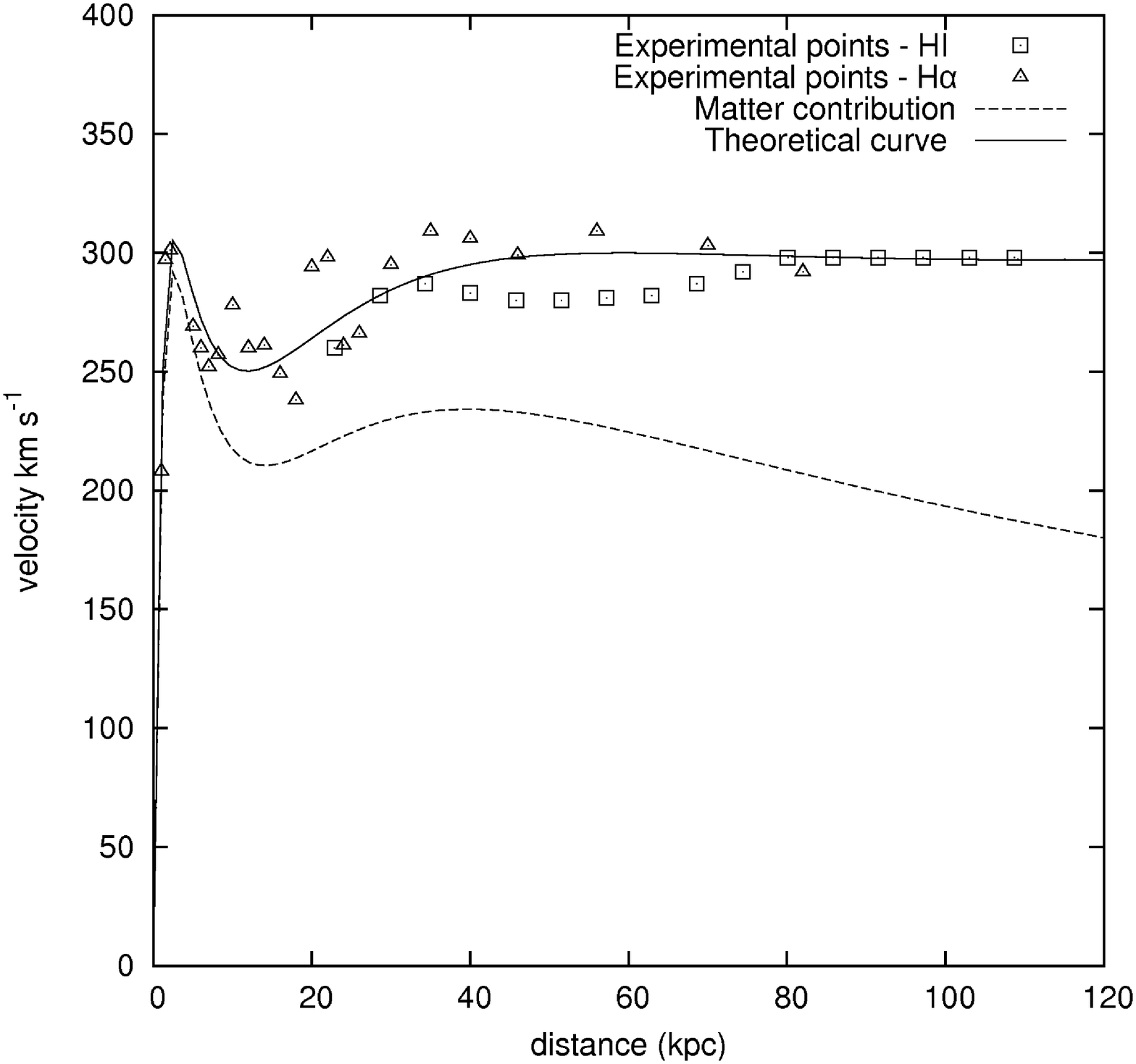}}
  \caption[Rotation curve for galaxy UGC 2885]{\label{fig3} Plot of
  the rotational velocity $v$ versus
 distance $r$ from the galactic center, for the
  galaxy UGC 2885. Squares ($\boxdot$) are the velocities determined from HI
  observations (see Roelfsema\&Allen \cite{roelfsema85}), 
  while triangles ($\triangle\kern
  -1.4ex \cdot \;$) are H$\alpha$ observations (see Rubin et al. 
\cite{rubin85}).  
  The solid line is the theoretical curve (see text), while the dashed
  line is the contribution due to the local matter. 
  }
\end{figure}

To make a comparison between the predictions given by
(\ref{eq:Fond2}), and the observed rotation curves of galaxies, one
should have available an expression for $V^{loc}(r)$, or, equivalently,
one should  assign the local distribution of matter. In the literature, there
are several papers which illustrate
different ways in which  the distribution of matter can be estimated, 
starting for
example from the measured distribution of luminosity of the galaxies
(see Toomre \cite{toomre}; Freeman \cite{freeman};
Carignan\&Freeman\cite{carignan};  Carignan et al. \cite{carignan88}).  
\begin{table}[!hb]
  \caption[Fit parameters]{\label{tab1} Values of total mass,
  mass--luminosity ratio and correlation length for the four galaxies
  considered.} 
  \centering 
  \begin{tabular}{c c c c }
    \hline
    \noalign{\smallskip}
      Galaxy &  Mass & Mass/Luminosity & Correlation length \\
      \hline
      NGC 3198 & 4.0 10$^{10}$ $M_{\sun}$ & \ \  4.6 
        $M_{\sun}/L_{\sun}$ & \quad  0.6 Kpc \\
      NGC 2403 & 3.5 10$^{10}$ $M_{\sun}$ & \ \  4.4 
        $M_{\sun}/L_{\sun}$ & \quad 0.8 Kpc \\
      NGC 2885 & 1.0 10$^{12}$ $M_{\sun}$ & \ \  2.1 
        $M_{\sun}/L_{\sun}$ & \quad 1.7 Kpc \\
      NGC 4725 & 1.1 10$^{11}$ $M_{\sun}$ & \ \  2.1 
        $M_{\sun}/L_{\sun}$ & \quad 3.1 Kpc \\
        \noalign{\smallskip}
        \hline
  \end{tabular}
\end{table}

We take instead the simpler path which consists in assuming a functional 
form for
$V^{loc}(r)$, with a minimal number of parameters (essentialy two,
total mass and radius of the galaxy) and then trying to determine these
parameters by a best fit with the
observed rotation curves. In particular we concentrate on the rotation
curves of the four 
galaxies NGC~3198, NGC~2403, UGC~2885 and NGC~4725. 

For what concerns the local potential $V^{loc}(r)$, 
in the literature one often makes use of potentials of the form 
$$
V^{loc}(r,z)= \frac {GM}{\sqrt{r^2 + \big( a+\sqrt{z^2+b^2}\big)^2 } }
\ ,
$$
first introduced by  Miyamoto\&Nagai (\cite{miyamoto}). Here 
$M$ is the total mass of the object,
to be understood in the following sense. 
First one introduces the matter density $\rho$ from the
Poisson equation $\Delta_2 V^{loc}=4\pi G\rho$, then one easily checks
that the integral
of $\rho$, over the whole space, is equal to  the parameter $M$.

For what concerns the other two parameters $a$ and $b$, one has first
of all that the ratio $b/a$ determines the flatness of the mass distribution 
$\rho$: for
$b/a>1$, the distribution is essentially spherical,
while it reduces to a singular disk if the ratio vanishes. The
parameter $a$ (at fixed ratio $b/a$) is then the lenght scale of the
distribution $\rho$.

So, for the galaxies NGC 3198 and NGC 4725 we take  the potential
$V^{loc}(r)$ in the the Miyamoto--Nagai form, supposing that the only
relevant contribution to the potential comes from the star disk. In
the case of the other two galaxies, i.e, NGC 2403 and UGC 2885, it
appears that there is also a non negligible contribution from the
inner part of the galaxy (the so called bulge), so that we take the
local potential as the sum of two Miyamoto--Nagay potentials, with
different parameters, i.e., we take
$$
  V^{loc}(r,z) = \frac {GM_1}{\sqrt{r^2 + \big( a_1+\sqrt{z^2+b_1^2}\big)^2
  } } + \frac {GM_2}{\sqrt{r^2 + \big( a_2 +\sqrt{z^2+b_2^2}\big)^2 } } \ ,
$$
the first term referring to the bulge, the second to the disk.
In this case, the total mass $M$ of the galaxy will be given by $M=M_1+M_2$.

Now, using the above expressions for the local potentials, we made a
best fit to the actual observed curve in order to determine both the
parameters of the potential and the correlation length $\lcor$. The
resulting rotation curves are reported as solid lines in 
Figures~\ref{fig1}--\ref{fig3}. The dashed lines  
are instead  the rotation curves which would be obtained if
the contribution of the distant matter were neglected. As one sees,
the contribution of far matter is essential to keep flat the
rotation curves at large distances. The values of the masses $M$ (in
solar units $M_{\bigodot}$), the mass
luminosity ratios (in solar mass over solar luminosity
$L_{\bigodot}$ units), and  the correlation lengths $\lcor$ in kiloparsecs
for the four galaxies are reported in Table~\ref{tab1}.

\begin{figure}
 \centering
\resizebox{\hsize}{!}{\includegraphics{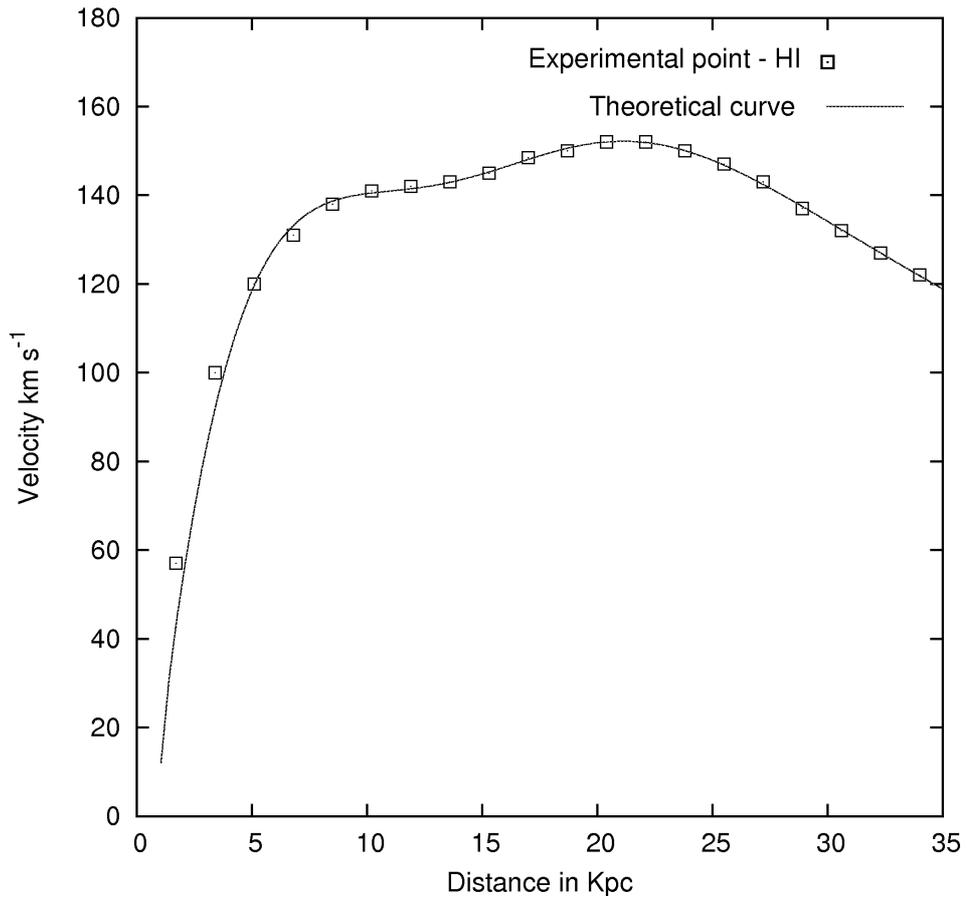}}
  \caption[Rotation curve for galaxy NGC 864]{\label{fig5} Plot of
  the rotational velocity $v$ versus distance $r$ from the galactic center, 
for the
  galaxy NGC 864. Squares ($\boxdot$) are the velocities determined from HI
  observations (see Espada et al. \cite{espada}), solid line is the
  theoretical curve (see text). }
\end{figure}

\begin{figure}
  \centering
\resizebox{\hsize}{!}{\includegraphics{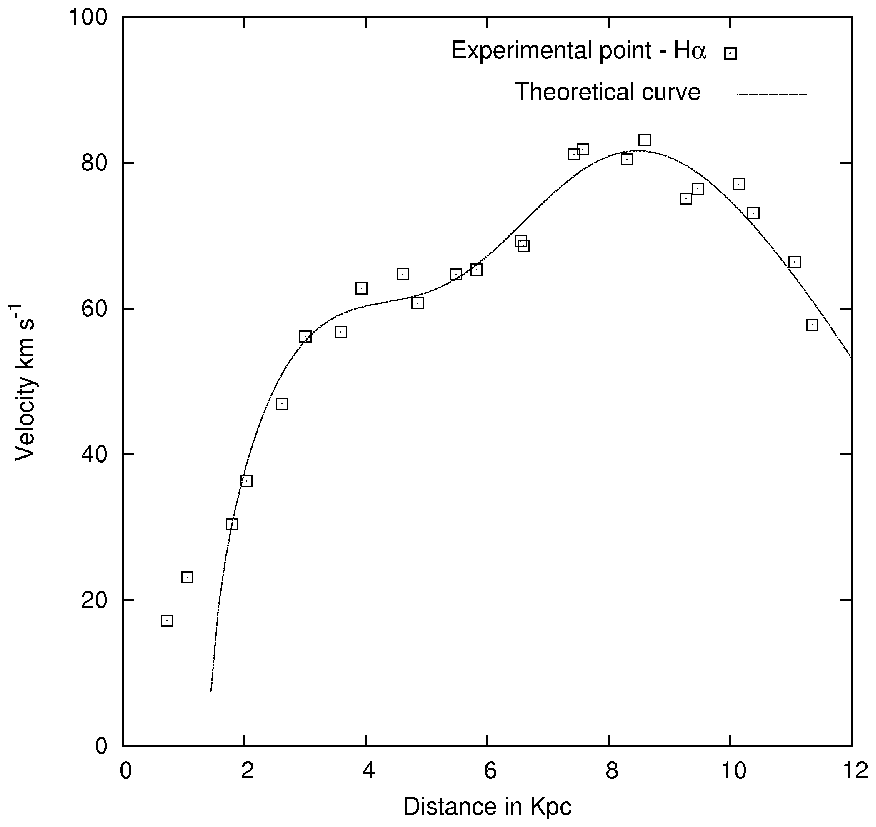}}
  \caption[Rotation curve for galaxy AGC 400848]{\label{fig6} Plot of
  the rotational velocity $v$ versus 
   distance $r$ from the galactic center, for the
  galaxy AGC 400848. Squares ($\boxdot$) are the velocities determined
  from H$\alpha$
  observations (see Catinella et al. \cite{palomar}), solid line is the
  theoretical curve (see text). 
  }
\end{figure}

At variance with the effect due to dark matter, the far
away matter could in principle also produce a \textsl{negative
  pressure}, because, as already explained,
the term $\partial_r V^{eff}$ in equation (\ref{eq:Fond2}) can be
positive or negative.
This would imply that the rotation curve decays, beyond the galaxy's
edge, faster than keplerian. By a quick search,
we found in literature at least two
cases of this behavior: the rotation curve of the galaxy NGC 864,
(see Espada et al. \cite{espada}) and that of galaxy 
AGC 400848 (see Catinella et al. \cite{palomar}). We are able to fit
also such curves, as shown in Figures~\ref{fig5} and \ref{fig6}, with
again a value of the correlation lenght $l$ of the order of 1 kpc. 

In any case, these  curves are shown here
only for the sake of illustration: in fact, first of
all the contribution of the local matter was computed with an ``ad
hoc'' potential without a deep confrontation with the data taken from
luminosity to see if they are compatible. 
More importantly, we did not investigate whether there are some perturbations
coming from known objects (such  as nearby galaxies), which could perturb the
velocity profile.

\section{Conclusions}

It was shown here that the faraway matter can exert
a force of such a  magnitude as to account for the observed
rotation velocity curves
in spiral galaxies, if the force is assumed to be decorrelated beyond
a sufficiently large distance, of the order of 1 kpc.  
In particular we fit the rotation curves of the galaxies NGC~3198,
NGC~2403, UGC~2885 and NGC~4725 without  any need to introduce dark
matter at all. 

We also showed that in principle faraway matter could act as a
\emph{negative pressure}, which would steepen the fall of
of the rotation curves (with respect to the keplerian decay), as
apparently observed for some galaxies. We considered, in a preliminary
way, two such  cases, for which our approach seems to apply, giving a
quite good fit. This subject is however left for a future work.

\subsection*{Acknowledgements}
 The author wants to thank prof. L.~Galgani for his many and  deep
 suggestions.

\appendix

\section{Computation of variances}

In this appendix we compute the average and the variance of  the terms
which appear in the equation of motion (\ref{eq:Fond1}). 
We recall that, from
the fourth of (\ref{eq:gcordcil}) of Section~3 and from the expression of
$h_{\mu\nu}$ given by \ref{eq:RetPot} of Section~2,  one has
\begin{equation}
\A =  - \frac {4G}{c^4} \sum_{j} M_j \frac {\big( 
  \dqj_x \cos\theta + \dqj_y \sin\theta \big)^2}{|\vqj - \vett x |}
  -\frac {2G}{c} \sum_{j} \frac { M_j}{|\vqj - \vett x |} \ ,
\end{equation}
where we are using cylindrical coordinates $(r,\theta,z)$, 
and the point $\vett x$ at which we are evaluating the field lies
on the galactic plane $z=0$. 

We have now to consider $\A_1 = \media{1}{\A}$, and in particular
$\langle \A_1 \rangle$. However, one obviously  has
$$ 
\langle \A_1 \rangle = \media{1}{\langle \A \rangle} = \langle \A
\rangle \ ,
$$
because $\langle \A \rangle$ is independent of the angle $\theta$,
due to the isotropy of the  $\vqj$ distribution. So we can take
$\theta=0$, and we obtain
\begin{eqnarray*}
     \langle \A \rangle &=& - \frac {4GH_0^2}{c^4} \langle  \sum_{j} M_j
     \frac { (\dqj_x)^2 }{|\vqj|} \rangle  - \frac{2G}{c^2}\langle
     \sum_{j} \frac {M_j}{|\vqj|} \rangle \\ 
     &=&  - \frac {4GH_0^2}{c^4}
     \langle \sum_{j} M_j  R_j \sin^2\phi_j\sin^2\theta_j \rangle - 
     \frac{2G}{c^2}\langle  \sum_{j} \frac {M_j}{|\vqj|} \rangle \ ,
\end{eqnarray*}
where $R_j=|\vqj|$ is the distance of the source $\vqj$ from the galaxy
center, $\theta_j$ is the angle between the source $\vqj$ and the galactic
plane, and $\phi_j$ a third angle on the galactic plane. We have
obviously omitted any term of  order $|\vett x|/R_j$. To compute
the average, we can take the continuum limit (as was done in paper
Carati et al. \cite{carati}), setting $M_j=\rho_{eff}\dif V$, so that one gets
\begin{eqnarray*}
    \langle &\A& \rangle = - \frac {2G\rho_{eff}}{c^4} \Big( 2H_0^2 \int
    R^3 \sin^2 \phi \sin^3 \theta \dif R\dif \theta\dif \phi + \\
    &~&c^2  \int
    R \sin\theta \dif R \dif \theta \dif\phi \Big) 
    = - \frac {2G}{c^4} \rho_{eff} H_0^2R_u^4\big( \frac \pi3 +\pi
    \big) =-\frac 12 \ ,
\end{eqnarray*}
where we have used the expression (8) of Carati et al. (\cite{carati}) for the
effective density $\rho_{eff}$. This gives 
the first of (\ref{eq:scarti}).

\vskip 1.ex
\noindent
We come now to the second of (\ref{eq:scarti}). In the first place one
has to consider that, in computing the derivative which appears in
the forces, one gets many terms, the largest of which
comes from the fact the
time at which we take the position of the sources is retarded,
i.e. is a function of $\vett x$. One has then 
\begin{eqnarray*}
    \partial_r \A &=&  - \frac {4G}{c^4} \sum_{j} M_j \frac {2\big( 
      \dqj_x \cos\theta + \dqj_y \sin\theta \big)\big( 
      \ddqj_x \cos\theta  + \ddqj_y \sin\theta \big) }{|\vqj - \vett x |} \\
    &-& \frac {2G}{c} \sum_{j} \frac { M_j}{|\vqj - \vett x |} \frac
{\partial_r |\vqj - \vett x |}{c} +   \mbox{smaller terms}\ ,
\end{eqnarray*}
where ``smaller terms'' has to be understood in the sense that they
give a smaller contribution to the mean and to the variance. Now one has
\begin{eqnarray*}
    \partial_r |\vqj - \vett x | &=& \partial_r \Big( \qj +r 
    \frac {\big( \dqj_x \cos\theta + \dqj_y \sin\theta \big)}{\vqj} +
    O\big((r/\qj)^2\big) \Big) \\
    &=&\frac {\big( 
      \dqj_x \cos\theta + \dqj_y \sin\theta \big)}{|\vqj|} + O(r/\qj) \ ,
\end{eqnarray*}
where we have used the fact that $\qj \gg r$. We get then, using 
Hubble's law, the following expression
\begin{equation}\label{eq:A-derA}
\partial_r \A =  - \frac {8GH^3}{c^5} \sum_{j} M_j \frac {2\big(
  \qj_x \cos\theta + \qj_y \sin\theta \big)^3 }{|\vqj|^2} \ ,
\end{equation}
where, from now on, we consider only the most relevant terms in the
expressions. Now, due to the rotational invariance of the
distribution, one gets immediately
$$
\langle \media{1}{\partial_r \A} \rangle = \media{1}{\langle
\partial_r \A \rangle} = 0 \ , 
$$
i.e. the first equality of the second line in (\ref{eq:scarti}). The estimate of
the variance $\sigma_{\A_2}$ is obtained in the following way.
By definition one has
$$
\sigma_{\A_2} = \Big\langle \frac 1{4\pi^2r^2}\oint\dif l_1 \oint \dif l_2
\partial_r A(\vett x) \partial_r A(\vett y) \Big\rangle  \ , 
$$
and using the hypothesis on the correlations one gets
$$
\sigma_{\A_2} = \frac 1{4\pi^2r^2}\oint\dif l_1 \oint \dif l_2
\Big\langle (\partial_r A)^2 \Big\rangle \exp\Big( - \frac {|\vett x
  -\vett y|}{\lcor} \Big)  \ . 
$$
Now, due to the rotational invariance of the distribution, the average
inside the integral does not depend on the angle $\theta$
and can be taken out from the integral. 
So one gets
$$
\sigma_{\A_2} = \frac {\Big\langle (\partial_r A)^2
  \Big\rangle}{4\pi^2r} \oint\dif l_1 \int_{0}^{2\pi} \dif \theta
 \exp\Big( - \frac {2r}{\lcor}\sin\theta/2 \Big) = \gamma_1 \frac
  {\lcor}{r} \Big\langle 
 (\partial_r A)^2 \Big\rangle  \ . 
$$
Now, the factor $\gamma_1$, which is essentially the value of the
inner integral, depends indeed on the ratio $r/\lcor$, but in the range
of interest is essentially equal to its asymptotic value $1/\pi$ (such
a value can be computed for example by applying Laplace method to the
inner integral). So we simply put
$$
\int_{0}^{2\pi} \dif \theta
 \exp\Big( - \frac {2r}{\lcor}\sin\theta/2 \Big)=\frac 1\pi \frac
 \lcor r  \ ,
$$
having stipulated that $\lcor/r$ is sufficiently large.

We are left with the computation of the variance $\langle
(\partial_r A)^2 \rangle$, which can be performed as follows. From
relation (\ref{eq:A-derA}), taking the square and using the
rotational invariance of the distribution to compute it for $\theta=0$,
one gets the following expression
$$
\langle (\partial_r \A)^2 \rangle =  \Big( \frac {8GH^3}{c^5}\Big)^2
 M^2 \sum_{j,k} 
 \frac {\big( \qj_x \big)^3\big( \qk_x\big)^3 }{|\vqj|^2
 |\vqk|^2}= \gamma_A \Big( \frac {8GH^3}{c^5}\Big)^2 M^2_{tot} R_u^2 \ ,
$$
with $M_{tot} =NM$, where $N$ the total number of galaxies, while
$\gamma_A$ is a numerical coefficient which depends on the 
distribution of the galaxies. If we choose the distribution as
discussed in Section~2, one can estimate it by numerical computations,
which give $\gamma_A\simeq 0.05$. 
Now, using  for $M_{tot}$ the expression
\begin{equation}\label{eq:massa}
M_{tot}= (4/3)\pi \rho_{eff} R_u^3 \ ,
\end{equation}
with $\rho_{eff}$ as given by (8) of Carati et al. (\cite{carati}), one
eventually finds 
$$
  \sigma_{\A_2} =  \frac {\gamma_A}{\pi}\frac {\lcor}{r} \left (
  \frac{H_0}{c} \right)^2  \ , 
$$
which is the second relation in the second line of (\ref{eq:scarti}). 

\vskip 1. ex
\noindent
Let us now come to the third and fourth lines of (\ref{eq:scarti}). The
average of $\B_1$ vanishes by 
symmetry, as can be seen from its expression
$$
\B_1 =   - \frac {4GH_0}{c^3}\media{1}{ \sum_{j} M_j \frac { 
  \qj_y \cos\theta - \qj_x \sin\theta }{|\vqj|} } \ ,
$$
where use was made of Hubble's law. For what concerns the variance
$\sigma_{\B_1}$ one has
\begin{eqnarray*}
    \sigma_{\B_1} &=&  \frac {1}{4\pi^2 r^2}\oint\dif l_1\oint \dif l_2
    \langle \B(\vett x)\B(\vett y)\rangle \\
    &=& \frac {\langle \nabla
      \B^2)\rangle}{2\pi} \int_0^{2\pi} \dif \theta \frac{\lcor^3}{2r\sin
      \theta/2} \Big( 1-e^{-2\frac {r}{\lcor}\sin \frac \theta2} \big( 1 +\frac
	    {r}{\lcor} \sin \frac \theta2 \big) \Big) \ ,
\end{eqnarray*}
where we used (\ref{eq:wk2}) of Appendix~B to estimate the correlation
of a function in terms of the correlations of its derivatives. Use was
also made of the isotropy of the distribution, so that $\langle \nabla
\B^2)\rangle$ turns out to be constant along the integration path. The
integral in the second line can be computed numerically, and with a
good accuracy (in the range of interest) 
one finds the value $2\pi \lcor^2 (\lcor/r)^{0.8}$ so that one has
\begin{equation}\label{eq:A1}
    \sigma_{\B_1} \simeq   \langle (\nabla
      \B)^2\rangle \lcor^2\left(\frac {\lcor}{r}\right)^{0.8} \ .  
\end{equation}
Exactly in the same way one can check that the average of $\B_2$
vanishes, while the variance $\sigma_{\B_2}$ is given by
\begin{equation}\label{eq:A2}
    \sigma_{\B_2} \simeq   \langle (\partial_r
      \B)^2\rangle \frac {\lcor}{\pi r} \ .  
\end{equation}
The expression of $\partial_r \B$, retaining only the largest terms,
is 
$$
\partial_r \B =-\frac {4G}{c^4} \sum_j M_j \frac {(\ddqj_y\cos\theta
  - \ddqj_x\sin\theta)(\qj_x\cos\theta + \qj_y\sin\theta)) 
}{|\vqj|^2} \ ,
$$
so that, using Hubble's law, one gets
$$
\langle (\partial_r \B)^2 \rangle = 
\Big(\frac {4GH_0^2M}{c^4} \Big)^2 \langle \, \sum_{j,k} \frac
    {\qj_y\qj_x\qk_x\qk_y}{|\vqj|^2|\vqk|^2 } \, \rangle
    = \gamma_B \Big(\frac {4GH_0^2M_{tot}}{c^4} \Big)^2 \ .
$$
Here, as before, one has $M_{tot}=NM$ where $N$ is
the number of galaxies, while
the numerical coefficient $\gamma_B$ depends on the distribution of
the galaxies; for our fractal model a numerical estimate gives
$\gamma_B\simeq 0.026 $. Again, inserting this
expression into relation (\ref{eq:A2}), and using for $M_{tot}$
the value (\ref{eq:massa}),   one gets the fourth
line of relation (\ref{eq:scarti}). The third line is obtained simply
by observing that
$$
    \langle(\nabla \B)^2\rangle = 3 \langle (\partial_r \B)^2\rangle\ .  
$$

\vskip 1.ex
\noindent
We come now to the fifth line of (\ref{eq:scarti}). One has 
$$
cg_{r0} =   - \frac {4G}{c^2}\sum_{j} M_j \frac { 
  \dqj_x \cos\theta + \dqj_y \sin\theta }{|\vqj-\vett x|}  \ ,
$$
so that one gets (forgetting as usual all small terms) the
relation
\begin{eqnarray*}
 c\partial_t g_{r0} &=& - \frac {4GH_0^2}{c^2}\sum_{j} M_j \frac { 
  \qj_y \cos\theta - \qj_x \sin\theta }{|\vqj-\vett x|} \\
  &=& \frac {4GH_0^2}{c^2}\sum_{j} M_j 
 \frac {\vqj_y\cdot\hat{\vett r}}{|\vqj-\vett x|}\ .
\end{eqnarray*}
From this expression it follows immediately that $\langle \partial_t
g_{r0} \rangle=0$ so that one has also  
$$
 \langle \C \rangle = \media{1}{ \langle c\partial_t g_{r0}\rangle} = 0\ , 
$$
which is the first relation of the last line in (\ref{eq:scarti}). The
computation of the variance $\sigma_{\C}^2$ is made in exactly the
same way as was done before for all other variances, obtaining
\begin{eqnarray*}
    &\sigma_{\C}^2&  =  \frac {c^2}{4\pi^2 r^2}\oint\dif l_1\oint \dif l_2
    \langle \partial_t g_{r0}(\vett x)\partial_t g_{r0}(\vett y)\rangle \\
    &=& \frac {c^2 \langle (\partial_t g_{r0})^2)\rangle}{2\pi} 
       \int_0^{2\pi} \dif \theta \exp\Big( - \frac
    {2r}{\lcor}\sin\theta/2 \Big)= \frac {c^2\lcor}{\pi r}\langle (\partial_t
    g_{r0})^2)\rangle\ .
\end{eqnarray*}
We are reduced to the computation of $\langle (\partial_t
g_{r0})^2)\rangle$,  and one gets
$$
\langle (\partial_t g_{r0})^2)\rangle = \gamma_C H_0^2  \ ,
$$ 
with the factor $\gamma_C \simeq 0.05$ for our fractal 
model. This numerical estimate was already done by Carati et
al. (\cite{carati}), in providing the estimate (11) of page four.
So also the fifth line of (\ref{eq:scarti}) is proven.

\vskip 1.ex
\noindent
We finally  come to the last relation  (\ref{eq:scarti}). One has
(again taking only the largest terms)
\begin{eqnarray*}
    \partial_t &\D& =   - \frac {4G}{c^4}\sum_{j} M_j \frac
    {\ddqj_x\dqj_x - \ddqj_y\dqj_y}{|\vqj-\vett x|} \sin2\theta \\
     &~& \qquad \qquad +  \frac {4G}{c^4}\sum_{j} 
    M_j \frac {\ddqj_x\dqj_y + \ddqj_y\dqj_x}{|\vqj-\vett x|} 
    \cos2\theta \\ 
     &=& \frac {4GH_0^3}{c^4}\sum_{j} M_j \frac
    {\big(\qj_x\big)^2 - \big(\qj_y\big)^2 }{|\vqj-\vett x|} \sin2\theta  
     +  M_j \frac {2\qj_x\dqj_y}{|\vqj-\vett x|} \cos2\theta \ ,
\end{eqnarray*}
which shows, in a very simple way, that one has  $\langle \partial_t \D
\rangle=0$. For what concerns the
variance $\sigma_{\D_1}^2$ of $\D_1$ the usual correlation argument
shows that
$$
    \sigma_{\D_1}^2 = \frac {\lcor}{\pi r}\langle (\partial_t\D)^2\rangle\ .
$$
As before, in the expression of $\langle
(\partial_t\D)^2\rangle$ one can put, for example, $\theta=0$, so that
one gets   
\begin{eqnarray*}
  \langle (\partial_t\D)^2\rangle
  &=& \left(\frac {4GH_0^3M}{c^4} \right)^2 \Big \langle \sum_{jk}  \frac
  {4 \qj_x\qj_y\qk_x \qk_y }{|\vqj||\vqk|} \Big \rangle \\
 &=&  \gamma_D \left(\frac {16GH_0^3M_{tot}}{c^4} \right) R_0^2  \ ,
\end{eqnarray*}
where $M_{tot}=NM$ is the total mass of the galaxies, $R_0=c/H_0$ is 
the radius of the
universe, $\gamma_D$ a numerical factor which depends on the 
distribution of the galaxies. 
For our fractal model,
a numerical estimate provides $\gamma_D\simeq 0.08 $. 
Now, using for $M_{tot}$ the value given by
(\ref{eq:massa}), one eventually finds
$$
  \langle (\partial_t\D)^2\rangle
    = \frac {\gamma_D}{4} H_0^2 \ ,
$$
which, inserted in the expression for $\sigma_{\D_1}^2$, gives the
second of the last line of (\ref{eq:scarti}).
This completes the proof.

\section{The Wiener--Khinchin theorem}\label{appendixB}
We deal here with the problem of how one
computes the autocorrelation $\langle f(\vett x) f(0) \rangle$ if the
autocorrelation 
$\langle \partial_{\vett x} f(\vett x) \partial_{\vett x} f(0) \rangle$
of the derivatives is known (see (\ref{eq:corr1}). We will show that if
one assumes
\begin{equation}\label{eq:wk1}
\big\langle \partial_{\vett x} f(\vett x) \partial_{\vett x} f(0) \big
   \rangle = \big\langle \big(\partial_{\vett x} f(0)\big)^2 \big \rangle
   \exp(-\frac {|\vett x|}{\lcor}) \ ,
\end{equation}
then one has
\begin{equation}\label{eq:wk2}
  \big\langle f(\vett x) f(0) \big\rangle =  \frac{\lcor^3}{|\vett x|} 
    \Big( 1-e^{-|\vett x|/\lcor} \big( 1 +\frac {|\vett x|}{2\lcor}\big) \Big) 
     \big\langle \big(\partial_{\vett x}
  f(0)\big)^2 \big\rangle \ .
\end{equation}
This gives formule (\ref{eq:corr2}), if we take for $f(x)$ the 
function $g_{\mu\nu}$.

To show this, we make use of the Wiener--Khinchin formula (see
Wiener \cite{wiener}; Khinchin \cite{kinchin})
\begin{equation}\label{eq:wk3}
  \langle f(\vett x) f(0) \rangle = \int_{\reali^3} |\hat f(\vett
  k)|^2 e^{i\vett k\cdot \vett x} \dif\vett k \ ,
\end{equation}
where $\hat f(\vett k)$ is the Fourier transform of $f(\vett
x)$. Now, the Fourier transform of $\partial_{\vett x} f(\vett x)$ is simply
given by $\vett k \hat f(\vett k)$, so that, once $\langle
\partial_{\vett x} f(\vett x) \partial_{\vett x} f(0) \rangle$ is
known, the spectrum $|\hat f(\vett k)|^2$ (and thus the correlation) is
easily found. 

Indeed, inverting the Wiener--Khinchin formula, the spectrum is given by
\begin{eqnarray*}
    &~&k^2 |\hat f (\vett k)|^2 = \frac 1{(2\pi)^3}\int_{\reali^3} 
    \big\langle \partial_{\vett x} f(\vett x) \partial_{\vett x} f(0) \big
    \rangle     e^{-i\vett k\cdot \vett x} \dif\vett x \\
    &=& \frac {\big\langle \big(\partial_{\vett x}
      f(0)\big)^2\big\rangle}{(2\pi)^3} \int_{\reali^3} \exp\Big(
    -\frac r\lcor\Big) \exp(-ikr\cos\theta)) r^2\sin\theta\dif
    r\dif\theta\dif\phi \ , 
\end{eqnarray*} 
where, in the second line, we used   polar spherical
coordinates $(r,\phi,\theta)$ in $\reali^3$ with the $z$--axis
parallel to $\vett k$. The integral is elementary, and one obtains
$$
  k^2 |\hat f (\vett k)|^2 = \frac {\big\langle \big(\partial_{\vett x}
  f(0)\big)^2\big\rangle}{2\pi^2} \frac {\lcor^3}{(1+\lcor^2 k^2)^2} \ ,
$$ 
which gives the spectrum
\begin{equation}\label{eq:wk4}
  |\hat f (\vett k)|^2 = \frac {\big\langle \big(\partial_{\vett x}
  f(0)\big)^2\big\rangle} {2\pi^2 k^2} \frac {\lcor^3}{(1+\lcor^2 k^2)^2} \ .
\end{equation} 
Now the correlation follows from the
Wiener-Khinchin formula quite
easily because, using spherical coordinates in the momentum space
$\vett k$ with the $z$--axis parallel to $\vett x$, and the expression
(\ref{eq:wk4}) just found for the spectrum, relation (\ref{eq:wk3}) gives
\begin{eqnarray*}
    &~&\big\langle f(\vett x) f(0) \big  \rangle  =  
    \frac {\big\langle \big(\partial_{\vett x}
    f(0)\big)^2\big\rangle}{2\pi^2} \int_{\reali^3} \frac
    {\lcor^3}{(1+\lcor^2 k^2)^2} \exp(i|\vett x|k\cos\theta) \\
    &~&\sin\theta \dif k\dif\theta\dif\phi 
    = \frac {\big\langle \big(\partial_{\vett x}
    f(0)\big)^2\big\rangle}{\pi} \int_{\reali} \frac{\sin(|\vett
    x|k)}{k} \frac {\dif k}{(1+\lcor^2 k^2)^2} \ .
\end{eqnarray*} 
The integral in the second line can be computed for example by using
the method of the residues, thus obtaining relation (\ref{eq:wk2}).

\end{document}